# A CYBER THREAT INTELLIGENCE MANAGEMENT PLATFORM FOR INDUSTRIAL ENVIRONMENTS

Alexandros Papanikolaou[1], Aggelos Alevizopoulos[1], Christos Ilioudis[2], Konstantinos Demertzis[3,*] Konstantinos Rantos[3]

[1]Innovative Secure Technologies P.C.;
a.papanikolaou@innosec.gr; a.alevizopoulos@innosec.gr
[2]International Hellenic University, Department of Information and Electronic Engineering;
iliou@ihu.gr
[3]International Hellenic University, Department of Computer Science;
kdemertzis@teiemt.gr, krantos@cs.ihu.gr

## ABSTRACT

*Developing intelligent, interoperable Cyber Threat Information (CTI) sharing technologies can help build strong defences against modern cyber threats. CTIs allow the community to share information about cybercriminals' threats and vulnerabilities and countermeasures to defend themselves or detect malicious activity. A crucial need for success is that the data connected to cyber risks be understandable, organized, and of good quality. The receiving parties may grasp its content and utilize it effectively. This article describes an innovative cyber threat intelligence management platform (CTIMP) for industrial environments, one of the Cyber-pi project's significant elements. The suggested architecture, in particular, uses cyber knowledge from trusted public sources and integrates it with relevant information from the organization's supervised infrastructure in an entirely interoperable and intelligent way. When combined with an advanced visualization mechanism and user interface, the services mentioned above provide administrators with the situational awareness they require while also allowing for extended cooperation, intelligent selection of advanced coping strategies, and a set of automated self-healing rules for dealing with threats.*

## KEYWORDS

*Cyber Threat Intelligent; Cyber Threat Information; Information Sharing; Industrial Environment; Cybersecurity;*

## 1. INTRODUCTION

The rapid development of new technologies in recent decades has significantly affected human societies and the existing economy [1]. The highly digitized and interconnected environment shapes the new Cyberspace [2] [3] providing new possibilities and opportunities for organizations to develop extroversion activities and actions [4] [5]. However, this new cyber-ecosystem faces a number of challenges such as cybercrime, advanced persistent threats and cyberattacks, resulting in a climate of uncertainty and instability that threatens expected growth and prosperity [6] [7] [8]. The emergence of a new generation of cyber threats highlights the need to modernize the way these challenges are addressed [9] [10] [11], bypassing the until recently tactics of organizations that relied on the passive use of main security appliances, like firewalls to protect their information and anti-malware solutions [12] [13] [14] [15]. The characteristic of the complexity of modern threats is that most successful attacks are perceived only during the subsequent forensics procedures [16] [17].

As it is easily understood, ensuring a successful defense requires complete control in all attempts to exploit the vulnerabilities of the system, as a successful attack could take advantage of the existence of a single vulnerability. In everyday life, many system administrators are unable to fix all the vulnerabilities of a system in time as limited experience, non-automation, increased workload, software dependencies, use of old systems and lack of timely availability of critical patches are their main brake. On the contrary, the automated, intelligent collection and correlation of suspicious actions taking place in a network, in the framework of a single strategy, which will take into account the latest digital threats, can help to take appropriate measures to deal immediately and finally shield an organization from cyber threats [18] [19] , even though the sharing of cyber-threat intelligence is a challenging process [20].

A turning point in this process is the use of Indicators of Compromise (IOCs) that support the security decision-making process [21]. IOCs include malware signature IDs, malicious IP addresses, malicious checksum (MD5) malware, and malicious URLs or domain names of Botnets, as well as patch fixes, good practices in control measures, access control policies or removing unnecessary services, and modifying firewall settings [22] [23] [24]. In other words, this is a huge repository of knowledge with proven defense techniques, which are strengthened daily by adding updates.

Taking into account the gap presented in the intelligent, efficient, and unified application of the knowledge available based on the IOCs, this work presents an innovative architecture for utilizing the specific knowledge based on interoperable and intelligent methods of modern computing. The proposed CTIMP is an advanced and adaptive system for monitoring and timely detection of security events that threaten an organization, incorporating advanced technologies of analysis, automated management, and execution of corrective actions, offering interactive real-time security interaction.

The rest of this paper is structured as follows: Section 2 presents the background of the research approach. Section 3 is allocated to the presentation of the proposed architecture and finally, section 4 concludes the research.

## 2. BACKGROUND

It is important to emphasize that in recent years the need for a cooperative response to security incidents has been highlighted and very significant progress has been made in this area [25]–[27]. But the constant evolution of the methods and technologies used in cybercrime, creates complete on platforms that do not implement real-time information mechanisms, as anything else can be considered obsolete. Also, extremely important feature is the interoperability that can in the effective collection, improvement, analysis and sharing of cyber-attack data. A typical example of such an application is the method of Modi et al. [28] where they propose a multilevel architecture for the thorough analysis of heterogeneous data through the interaction between its interleaves. Although the approach accepts information from open-source data streams, it is considered to be completely dependent on in-house analysis platforms, which significantly limits the generalization that should be provided in such cases. Mantis presents a different approach that incorporates information on cyber threats using different standards [29]. It is an intelligent platform that allows threat data to be correlated through an innovative agnostic similarity algorithm. This methodology allows security analysts to correlate patterns that are shared between seemingly unrelated attacks, which adds serious complexity to the system, dramatically increasing the need for computing resources. Finally, Sengupta et al. [30] proposed a fairly sophisticated but very sophisticated method for extracting an optimal modeling technique for Advanced Persistent Threat attacks in a cloud computing environment. The approach is based on game theory, where the processes of dealing with an event are modeled by optimizing the cost of security countermeasures.

# 3. PROPOSED ARCHITECTURE

The proposed CTIMP, while following the practices of Integrated Security Information and Event Management (SIEM) [31], goes one step further by offering a personalized security solution that combines multiple control mechanisms and corresponding digital security technologies for modern computing systems and networks. Essentially, through a sophisticated collaborative framework, it is able to identify an organization's digital risks and threats, meeting the ongoing needs of securing the valuable information it manages, by offering security services and crisis remedies.

Specifically, CTIMP offers a central point of analysis, alert, compliance, and reporting, responding to the changing organizational structures of a multidimensional modern organization. It focuses primarily on meeting the critical information infrastructure needs of each organization, providing a variety of intelligent mechanisms for monitoring data integrity, reporting new threats, detecting and recording security incidents, and responding immediately to automated processes.

Taking a more detailed approach to the proposed architecture, the system initially focuses on the timely detection of events using automated, detailed log analysis. Any alerts or events are displayed on the system administrator's visualization console. This interface offers a timely and valid simultaneous analysis of a very large number of security incidents of the supervised business network while minimizing the possibility of drawing incorrect conclusions.

Updating and upgrading CTIMP predictability is based on gathering cyber-threat information from trusted open access sources such as ready-made IOCs set up by security experts, etc. which are filtered and correlated for the sole purpose of supporting infrastructure. This adaptation to the requirements of the organization's business operations and information systems is achieved through component mapping, which records and distributes to other CTIMP subsystems the specific features of the available node topology, software, and services that may be targets or sources of malicious action. Comparison of cyber-threats with the characteristics of the organization results in an adapted cyber-cognition in the STIX 2.x [32] standard, which is submitted for analysis to the privacy policy production subsystem, which provides algorithms for modulating and exporting SIGMA rules, which are automatically integrated into all the active rules of the case analysis mechanism. In addition, the proposed architecture includes an intelligent automated threat and attack mechanism, which provides an expanded set of self-healing commands, which examine the level of compliance and align it with current security policies.

Another very important feature is that the proposed architecture is assisted in each phase of its operation by the visualization and interface subsystem, which aims to easily represent the appropriate information, to allow analysts to detect and take immediate action in various events. Finally, it should be noted that the design of preventive countermeasures offered by CTIMP, concerns only the identification of specific threats that may affect the organization and the setting of general priorities, establishing intelligent decision-making or adaptation mechanisms, which ensure the smooth its operation.

A brief overview of CTIMP subsystems and functions is shown in Figure 1 below.

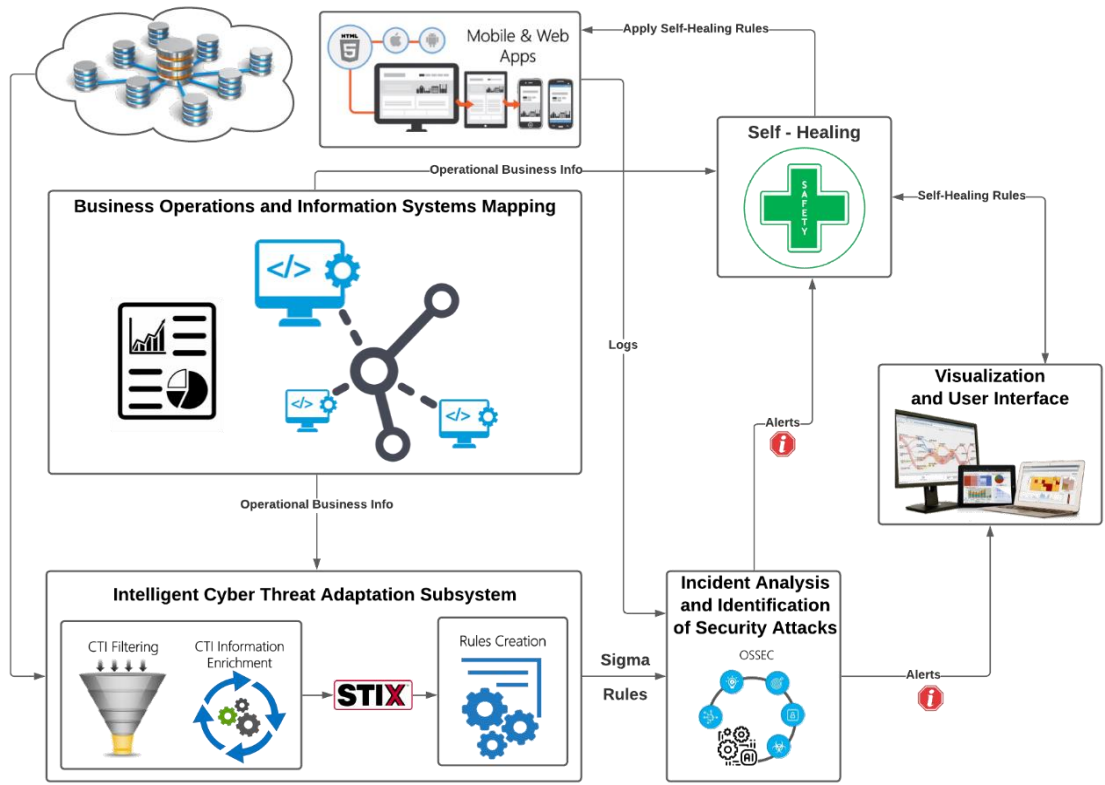

Figure 1: The proposed Cyber Threat Intelligent Information Sharing Architecture (CTIMP)

Below is a detailed presentation of the subsystems and mechanisms that frame the CTIMP architecture.

### 3.1 Incident analysis and identification of security attacks
The analysis of incidents and the intelligent identification of security incidents such as attacks are achieved in CTIMP by the following systems.

#### 3.1.1 OSSEC HIDS
The OSSEC Host-based Intrusion Detection System [33] conducts both application and system-level audits to investigate the integrity of supervised information infrastructure files, using threat detection methods based on signatures and statistical anomalies. It can be configured to collect events from devices on which the use of agents is not feasible, while also having a set of rules for monitoring and analyzing specialized security incidents, for which it can generate corresponding alerts.

#### 3.1.2 Decoders
The logs of the target environment are examined using default and custom decoders, which have parameters that are compared with the content of the logs for event detection. Any correspondences are routed for control by the set of available rules that implement the respective security policies, based on which the notifications of the incidents under consideration are produced.

### 3.2 Intelligent use of CTI
The Intelligent Cyber Threat Adaptation Subsystem collects and analyzes available information gathered from Cyber Threat Information (CTI) Sources. This information is compared with the results of the analysis of the data by the supervised Information System and the threats to the target network and information infrastructure are documented. The intelligent cybernetics aggregation mechanism is regularly upgraded with IOCs and RSS feeds, collected by the Malware Information Sharing and Threat Intelligence Sharing Platform (MISP) [34], from a wide range of trusted sources. The accuracy of the information is adjusted to the criteria, concerning the organization's systems, with a filtering process in order to maintain the relevant information and then to compare it with the specific characteristics of the supervised information infrastructure. Features include devices, services, objects IDs, IP addresses, geolocation information, and dependencies. The result is the production of custom knowledge in the form of STIX 2.x files, which correspond to the fields that are required to draft SIGMA rules.

### 3.3 Business operations and information systems mapping

The mapping mechanism describes the specific characteristics of the organization and the technological environment is displayed on the system's console. The latter provides a useful for manually recording business assets like nodes, communications, software, and network services. In particular, the Dependency Mapper utility implements the basic processes of the mechanism. The depmapper has a graphical data management interface to represent a deployment model. The adaptation of the depmapper to the mapping mechanism helps to extend the functionality of exporting graphs to image files (jpg, png) and data exchange (JSON). In practice, the production of image files supports collaboration between stakeholders, as well as the use of images between different third-party applications. Furthermore, the production of JSON-type files allows the insertion of graphs in the depmapper, for the purpose of their subsequent processing.

More specifically, the layout model illustrates the components, hardware components (nodes), and the links between them. The depmapper has the additional functions of grouping nodes, adding tags, and defining multiple descriptions separately for each node. The descriptions contain information about the object of operation, the available software services, the geographical location, the structural and procedural dependencies, the data dependencies, as well as the risk level. The level of risk is determined according to the criticality or sensitivity of the data and services available. The collection of this information takes place within the organization, by completing the user questionnaires. The information is collected by an analyzer and entered into the depmapper to represent graphs and text in JSON format, for use in other subsystems.

### 3.4 Visualization and user interface

This subsystem is divided into two subcategories, which include data visualization and user interface. Visualization of security data collected from various log sources refers to the creation of diagrams, graphs, and similar visual material. The visualization mechanism provides organization and classification of the data structure. A notice indicates that immediate action is required or is merely informative. Categorizing the different types of alerts is useful and makes it possible to develop a monitoring strategy [35]. Notifications are categorized according to the level of completion of the relevant research, conducted by managers and analysts.

An administrator is able to edit the content of an alert immediately, either later, or outsource its processing to an analyst, using ticketing requests. Consequently, the status of notifications is characterized as new, ongoing, and complete. In addition, the visualization mechanism provides all the data in a user-friendly environment. The user interface mechanism offers the possibility of approving or activating the system self-healing procedures. However, it is a fully customizable environment, which allows the rapid collection of critical information and immediate response to incidents. Likewise, it allows the viewing and analysis of critical statistics and the viewing of history. The interface supports web environments to make services available on mobile devices and web browsers.

### 3.5 Self-Healing Policies

Self-Healing Policies derive from the decision analysis and prioritization processes of the organization and are recorded in a clear and interoperable manner in the system database. The database consists of Threats, Policies, and Self-Healing Rules [36]. The Threat panel contains the fields of the threat id, the threat type, and the threat group. Self-healing commands are stored in a technical Command Line Interface (CLI) format, so that they can be understood by machines, as well as in a general format, readable by humans. The self-healing policy also includes the entries of the CLI commands, which concern the central nodes. CLI commands are properly synthesized to run on devices located at the ends of the network, including routers, switches, firewalls, agents, and AV software. Preventing a threat can be achieved by stopping the flow of network traffic in a timely manner, or by making inaccessible a device involved in the attack.

In particular, Self-Healing commands include three execution options, customizable through the system's console:
1. Inform the administrator about the actions to be taken in order to prevent a threat or mitigate the risk (recommendations).
2. Execution following the administrator's approval.
3. Automated execution, provided that the administrator has selected the specific configuration.

This subsystem receives data from the OSSEC system and from the Business operations and information systems mapping module of the organization, through control command flows. At the same time, it maintains two-way communication with the Visualization and User Interface subsystem, through data streams. The self-healing instructions are presented to the administrator and approval is required to execute a command. The administrator's decision is then forwarded to the self-healing subsystem. If the administrator approves the action, the command is executed immediately. If the action is rejected, then the self-healing command is given as a recommendation to the administrator. The set of requests and responses of the above communication is conducted asynchronously.

The Decision Engine, part of the Self-Healing module, determines the policy to be applied when an incident is detected. The procedure involves executing a command if the Threat Type field corresponds to a value in the Policies table. If the Threat Type field does not provide a value, then the more generic Threat Group field is checked instead. The event is then forwarded to the Visualization and User Interface subsystem and the relevant breach notifications are presented. In most cases the Self-Healing rules are applied remotely to the nodes, using the Secure Shell (SSH) protocol while the details on how to execute the alternative commands are recorded in a log file.

CTIMP exhibits the characteristics of complete network surveillance and effectively extends the settlement of security issues to other levels (systems, services). The technologies it incorporates help reduce the complexity of the methods used in today's attacks, by setting up specialized security software, with its main functions satisfying the need for regular checks to identify threats, update security policies, and maintain the organisation's security posture to an acceptable level.

## 4. CONCLUSIONS

An innovative architectural standardization of how to intelligently manage and deal with advanced cyber threats was presented in this paper. The proposed CTIMP in a fully interoperable and intelligent way, collaboratively utilizing cyber-knowledge generated on a daily basis by cyber-threat managers around the world, ensures high levels of security of an organization's supervised information infrastructure. Its design is based on standards that are able to maintain secure communication with reliable sources of information and receive regular updates on existing and emerging threats. The updates in question are initially optimized based on the respective needs and priorities of the information infrastructure of the supported organization. They are then transformed into automation rules that align the operating systems of the information systems and are finally applied in practice by implementing notifications to the system administrators for immediate action. It is an excellent mechanism for monitoring and timely detection of events in real-time, which significantly enhances the levels of active cyber security of an organization.

The most important task for the evolution of the proposed system is initially the process of finding solutions for the comparison of logs and security policies for their convergence in shorter times. Also, the strengthening of CTIMP with more advanced anomaly detection techniques which will take into account most of the operational parameters of the organization such as task scheduling, local events, technical upgrades or system adaptations, etc., would be a significant improvement. In addition, the system's structure should be examined to see how it might be utilized with data transformation methods, so that intelligent processes can discover the optimum techniques to represent various types of structured or unstructured data to provide self-healing rules. Finally, the CTI2SA's significant future growth must be based on interpretive models. These models may describe the decision process by defining individual predictions using approaches such as Shapley values and feature significance. The purpose of model interpretation is to extract human-comprehensible terminology for models' functioning mechanisms in researching adversarial movements and defenses.

# FUNDING

Co-financed by the European Regional Development Fund of the European Union and Greek national funds through the Operational Program Competitiveness, Entrepreneurship and Innovation, under the call RESEARCH – CREATE - INNOVATE (project code: T2EDK-01469)

# REFERENCES

[1] A. Al-Fuqaha, M. Guizani, M. Mohammadi, M. Aledhari, and M. Ayyash, "Internet of Things: A Survey on Enabling Technologies, Protocols, and Applications," *IEEE Commun. Surv. Tutor.*, vol. 17, no. 4, pp. 2347–2376, 2015, doi: 10.1109/COMST.2015.2444095.

[2] E. Harjula, A. Artemenko, and S. Forsström, "Edge Computing for Industrial IoT: Challenges and Solutions," in *Wireless Networks and Industrial IoT: Applications, Challenges and Enablers*, N. H. Mahmood, N. Marchenko, M. Gidlund, and P. Popovski, Eds. Cham: Springer International Publishing, 2021, pp. 225–240. doi: 10.1007/978-3-030-51473-0_12.

[3] M. O. Al Enany, H. M. Harb, and G. Attiya, "A Comparative analysis of MQTT and IoT application protocols," in *2021 International Conference on Electronic Engineering (ICEEM)*, Jul. 2021, pp. 1–6. doi: 10.1109/ICEEM52022.2021.9480384.

[4] A. Banafa, "2 The Industrial Internet of Things (IIoT): Challenges, Requirements and Benefits," in *Secure and Smart Internet of Things (IoT): Using Blockchain and AI*, River Publishers, 2018, pp. 7–12. Accessed: Jan. 19, 2021. [Online]. Available: https://ieeexplore.ieee.org/document/9226906

[5] M. Boubekeur, "Industrial applications for cyber-physical systems," in *2017 First International Conference on Embedded Distributed Systems (EDiS)*, Dec. 2017, pp. 59–59. doi: 10.1109/EDIS.2017.8284020.

[6] H. Chen, M. Hu, H. Yan, and P. Yu, "Research on Industrial Internet of Things Security Architecture and Protection Strategy," in *2019 International Conference on Virtual Reality and Intelligent Systems (ICVRIS)*, Sep. 2019, pp. 365–368. doi: 10.1109/ICVRIS.2019.00095.

[7] H. Geng, "THE INDUSTRIAL INTERNET OF THINGS (IIoT)," in *Internet of Things and Data Analytics Handbook*, Wiley, 2017, pp. 41–81. doi: 10.1002/9781119173601.ch3.

[8] M. J. Farooq and Q. Zhu, "IoT Supply Chain Security: Overview, Challenges, and the Road Ahead," *ArXiv190807828 Cs*, Jul. 2019, Accessed: Jan. 19, 2021. [Online]. Available: http://arxiv.org/abs/1908.07828

[9] K. Dawood, "An overview of renewable energy and challenges of integrating renewable energy in a smart grid system in Turkey," in *2020 International Conference on Electrical Engineering (ICEE)*, Sep. 2020, pp. 1–6. doi: 10.1109/ICEE49691.2020.9249780.

[10] W. Z. Khan, M. H. Rehman, H. M. Zangoti, M. K. Afzal, N. Armi, and K. Salah, "Industrial internet of things: Recent advances, enabling technologies and open challenges," *Comput. Electr. Eng.*, vol. 81, p. 106522, Jan. 2020, doi: 10.1016/j.compeleceng.2019.106522.

[11]   S. Rouhani and R. Deters, "Blockchain based access control systems: State of the art and challenges," *IEEEWICACM Int. Conf. Web Intell.*, pp. 423–428, Oct. 2019, doi: 10.1145/3350546.3352561.

[12]   K. R. Choo, S. Gritzalis, and J. H. Park, "Cryptographic Solutions for Industrial Internet-of-Things: Research Challenges and Opportunities," *IEEE Trans. Ind. Inform.*, vol. 14, no. 8, pp. 3567–3569, Aug. 2018, doi: 10.1109/TII.2018.2841049.

[13]   V. S. Mahalle and A. K. Shahade, "Enhancing the data security in Cloud by implementing hybrid (Rsa amp; Aes) encryption algorithm," in *2014 International Conference on Power, Automation and Communication (INPAC)*, Oct. 2014, pp. 146–149. doi: 10.1109/INPAC.2014.6981152.

[14]   K. Demertzis, K. Rantos, and G. Drosatos, "A Dynamic Intelligent Policies Analysis Mechanism for Personal Data Processing in the IoT Ecosystem," *Big Data Cogn. Comput.*, vol. 4, no. 2, p. 9, Jun. 2020, doi: 10.3390/bdcc4020009.

[15]   K. Demertzis, N. Tziritas, P. Kikiras, S. L. Sanchez, and L. Iliadis, "The Next Generation Cognitive Security Operations Center: Adaptive Analytic Lambda Architecture for Efficient Defense against Adversarial Attacks," *Big Data Cogn. Comput.*, vol. 3, no. 1, p. 6, Mar. 2019, doi: 10.3390/bdcc3010006.

[16]   H. Majed, H. N. Noura, and A. Chehab, "Overview of Digital Forensics and Anti-Forensics Techniques," in *2020 8th International Symposium on Digital Forensics and Security (ISDFS)*, Jun. 2020, pp. 1–5. doi: 10.1109/ISDFS49300.2020.9116399.

[17]   M. Stoyanova, Y. Nikoloudakis, S. Panagiotakis, E. Pallis, and E. K. Markakis, "A Survey on the Internet of Things (IoT) Forensics: Challenges, Approaches, and Open Issues," *IEEE Commun. Surv. Tutor.*, vol. 22, no. 2, pp. 1191–1221, 2020, doi: 10.1109/COMST.2019.2962586.

[18]   K. Rantos, G. Drosatos, K. Demertzis, C. Ilioudis, A. Papanikolaou, and A. Kritsas, "ADvoCATE: A Consent Management Platform for Personal Data Processing in the IoT Using Blockchain Technology," in *Innovative Security Solutions for Information Technology and Communications*, Cham, 2019, pp. 300–313. doi: 10.1007/978-3-030-12942-2_23.

[19]   S. Choi, J.-H. Yun, and S.-K. Kim, "A Comparison of ICS Datasets for Security Research Based on Attack Paths," in *Critical Information Infrastructures Security*, Cham, 2019, pp. 154–166. doi: 10.1007/978-3-030-05849-4_12.

[20]   K. Rantos, A. Spyros, A. Papanikolaou, A. Kritsas, C. Ilioudis, and V. Katos, "Interoperability Challenges in the Cybersecurity Information Sharing Ecosystem," *Computers*, vol. 9, no. 1, p. 18, Mar. 2020, doi: 10.3390/computers9010018.

[21]   D. Rhoades, "Machine actionable indicators of compromise," in *2014 International Carnahan Conference on Security Technology (ICCST)*, Oct. 2014, pp. 1–5. doi: 10.1109/CCST.2014.6987016.

[22]   B. Akram and D. Ogi, "The Making of Indicator of Compromise using Malware Reverse Engineering Techniques," in *2020 International Conference on ICT for Smart Society (ICISS)*, Nov. 2020, vol. CFP2013V-ART, pp. 1–6. doi: 10.1109/ICISS50791.2020.9307581.

[23]   V. Atluri and J. Horne, "A Machine Learning based Threat Intelligence Framework for Industrial Control System Network Traffic Indicators of Compromise," in *SoutheastCon 2021*, Mar. 2021, pp. 1–5. doi: 10.1109/SoutheastCon45413.2021.9401809.

[24]   M. Verma, P. Kumarguru, S. Brata Deb, and A. Gupta, "Analysing Indicator of Compromises for Ransomware: Leveraging IOCs with Machine Learning Techniques," in *2018 IEEE International Conference on Intelligence and Security Informatics (ISI)*, Nov. 2018, pp. 154–159. doi: 10.1109/ISI.2018.8587409.

[25]   Y. Gao, X. LI, H. PENG, B. Fang, and P. Yu, "HinCTI: A Cyber Threat Intelligence Modeling and Identification System Based on Heterogeneous Information Network," *IEEE Trans. Knowl. Data Eng.*, pp. 1–1, 2020, doi: 10.1109/TKDE.2020.2987019.

[26]   H. Zhao, Q. Yao, J. Li, Y. Song, and D. L. Lee, "Meta-Graph Based Recommendation Fusion over Heterogeneous Information Networks," in *Proceedings of the 23rd ACM SIGKDD International Conference on Knowledge Discovery and Data Mining*, New York, NY, USA, Aug. 2017, pp. 635–644. doi: 10.1145/3097983.3098063.

[27]   X. Liao, K. Yuan, X. Wang, Z. Li, L. Xing, and R. Beyah, "Acing the IOC Game: Toward Automatic Discovery and Analysis of Open-Source Cyber Threat Intelligence," in *Proceedings of the 2016 ACM SIGSAC Conference on Computer and Communications Security*, New York, NY, USA, Oct. 2016, pp. 755–766. doi: 10.1145/2976749.2978315.

[28] A. Modi *et al.*, "Towards Automated Threat Intelligence Fusion," in *2016 IEEE 2nd International Conference on Collaboration and Internet Computing (CIC)*, Nov. 2016, pp. 408–416. doi: 10.1109/CIC.2016.060.

[29] H. Gascon, B. Grobauer, T. Schreck, L. Rist, D. Arp, and K. Rieck, "Mining Attributed Graphs for Threat Intelligence," in *Proceedings of the Seventh ACM on Conference on Data and Application Security and Privacy*, New York, NY, USA, Mar. 2017, pp. 15–22. doi: 10.1145/3029806.3029811.

[30] S. Sengupta, A. Chowdhary, D. Huang, and S. Kambhampati, "General Sum Markov Games for Strategic Detection of Advanced Persistent Threats Using Moving Target Defense in Cloud Networks," in *Decision and Game Theory for Security*, Cham, 2019, pp. 492–512. doi: 10.1007/978-3-030-32430-8_29.

[31] S. Bhatt, P. K. Manadhata, and L. Zomlot, "The Operational Role of Security Information and Event Management Systems," *IEEE Secur. Priv.*, vol. 12, no. 5, pp. 35–41, Sep. 2014, doi: 10.1109/MSP.2014.103.

[32] "Introduction to STIX." https://oasis-open.github.io/cti-documentation/stix/intro.html (accessed Oct. 14, 2021).

[33] "OSSEC - World's Most Widely Used Host Intrusion Detection System - HIDS," *OSSEC*. https://www.ossec.net/ (accessed Oct. 14, 2021).

[34] "MISP - Open Source Threat Intelligence Platform & Open Standards For Threat Information Sharing (formely known as Malware Information Sharing Platform)." https://www.misp-project.org/ (accessed Oct. 14, 2021).

[35] Y. Yang *et al.*, "Dark web forum correlation analysis research," in *2019 IEEE 8th Joint International Information Technology and Artificial Intelligence Conference (ITAIC)*, May 2019, pp. 1216–1220. doi: 10.1109/ITAIC.2019.8785760.

[36] A. Spyros, K. Rantos, A. Papanikolaou, and C. Ilioudis, "An Innovative Self-Healing Approach with STIX Data Utilisation:," in *Proceedings of the 17th International Joint Conference on e-Business and Telecommunications*, Lieusaint - Paris, France, 2020, pp. 645–651. doi: 10.5220/0009893306450651.